\documentclass[11pt]{article}             
\title{Radon transform and kinetic equations in tomographic representation}
\author{V.~N.~Chernega$^{\dag}$, V.~I.~Man'ko$^{\dag\dag}$, B.~I.~Sadovnikov$^{\dag}$\\\\
\\$^{\dag}$Faculty of Physics, M.V. Lomonosov Moscow State University\\
Vorob'evy Gory, Moscow 119992, Russia
        \\$^{\dag\dag}$ P.N.~Lebedev Physical Institute, Russian Academy of Sciences\\
        Leninskii Prospect, 53, Moscow 119991, Russia \\
        \\Emails: vchernega@gmail.com, manko@sci.lebedev.ru, b.i.sadovnikov@gmail.com}

\begin{document}

\maketitle

\begin{abstract}
Statistical properties of classical random process are considered
in tomographic representation. The Radon integral transform is
used to construct the tomographic form of kinetic equations.
Relation of probability density on phase space for classical
systems with tomographic probability distributions is elucidated.
Examples of simple kinetic equations like Liouville equations for
one and many particles are studied in detail.
\end{abstract}

\section{Introduction}
The kinetic equations describing the behavior of classical system
can be obtained using procedure suggested by Bogolyubov
\cite{Bog1} where the chain of connected equations corresponding
to $N$ and $N+1$ particles in the system was studied (see also
\cite{BogSad}). The simplest Liouville equation for $N+1$
particles probability density on phase-space provided the more
complicated equation for $N$ particles probability density after
averaging over $(N+1)$th particle degree of freedom. The
appropriate reduction procedure of the Bogolyubov chain of
equations gives other kinetic equations, for example Boltzman
equation with the collision term. The probability density on
phase-space $f(\vec{q},\vec{p},t)\equiv
f(q_1,q_2,...,q_N,p_1,p_2,...,p_N,t)$ depends on positions and
momenta of the particles. Recently it was suggested to use for
description of quantum \cite{Mancini96} and classical
\cite{OlgaJRLR97} states the tomographic probability distribution
function (state tomogram). For classical particles the tomogram is
Radon transform \cite{Radon} of the probability density
$f(\vec{q},\vec{p},t)$ on the phase-space. The advantage of the
tomographic probability representation is connected with the fact
that in this representation both classical and quantum states are
described by the identical objects-tomographic probability
distributions. The physical meaning of the tomographic probability
distribution is the following one. For one particle it is
distribution of the particle position $X$ but this position in
measured is the reference frame in the system phase-space which
rotated by angel $\theta$ and before the rotation the axes
$q\rightarrow sq$, $p\rightarrow s^{-1}p$ where $s$ it the scaling
parameter. Then the scaled and rotated frame can be parameterized
by two real parameters $\mu=s\cos\theta$ and
$\nu=s^{-1}\sin\theta$. For one particle the tomogram is the
probability distribution function $w(X,\mu,\nu)$ of random
position $X$ and two real reference-frame parameters $\mu$, $\nu$.
For $N$ particles the tomogram is the function
$w(\vec{X},\vec{\mu},\vec{\nu})$ where $\vec{X}$ is vector with
particles positions $X_j,$, $j=1,2,...,N$ and $\vec{\mu}$ and
$\vec{\nu}$ vector components $\mu_j=s_j\cos\theta_j$ and
$v_j=s_j^{-1}\sin\theta_j$ describe the scaling and rotation
parameters of reference-frame axes associated with $j$th degree of
freedom. The aim of this work is the study of classical kinetic
equations in tomographic probability representation. We focuse on
the Bogolyubov chain written for the function
$f(\vec{q},\vec{p},t)$ and apply the Radon transform to write the
chain of kinetic equations. On the other hand the consideration
can be extended to quantum kinetic equations since in the
tomographic picture the connection of tomograms for $N+1$ and $N$
particles is identical for classical and for quantum domains. This
fact corresponds to analogous relations for Wigner functions
\cite{Wigner32} in quantum domain but the Wigner function can take
negative values and it means that the function is not probability
distribution function. Some aspects of the tomographic description
of classical and quantum states were considered in
\cite{Pilavets},\cite{CherManko1},\cite{CherManko2},\cite{CherManko3},\cite{CherManko4},\cite{CherManko5},\cite{CherManko6},\cite{repino}.
The paper is organized as follows. In Sec.2 we review the approach
with the tomographic Radon map of classical probability
distribution both for one particle and for $N$ particles. In Sec.3
we obtain the Liouville equation in the tomographic probability
representation. In Sec.4 we review the procedure with averaging
the Liouville equation used in Bogolyubov chain approach. Also we
consider the Radon representation. In Sec.5 the perspectives and
conclusions are given.

\section{Probability distribution and tomograms}
Let us consider a particle with fluctuating momentum $p$ and
position $q$. The state of such particle is associated with
probability distributions $f(q,p)$ on the particle phase space.
The probability distribution is nonnegative and normalized
function, i.e.
\begin{eqnarray}\label{eq.1}
f(q,p)\geq 0
\end{eqnarray}
and
\begin{eqnarray}\label{eq.2}
\int f(q,p)dqdp=1.
\end{eqnarray}
The function $f(q,p)$ can be considered also as generalized
function, e.g. Dirac delta-function. For many particles the state
of the system with fluctuations is described by the probability
distributions $f(\vec{q},\vec{p})$ where
$\vec{q}=(q_1,q_2,..,q_N)$, $\vec{p}=(p_1,p_2,..,p_N)$ but this
function is nonnegative and normalized as in one-dimensional case.
The Radon transform of the probability distribution function
$f(q,p)$ is gives by the formula
\begin{eqnarray}\label{eq.3}
w(X,\mu,\nu)=\int f(q,p)\delta(X-\mu q-\nu p)dqdp
\end{eqnarray}
where the arguments $X,\mu,\nu$ are real numbers. The function
$w(X,\mu,\nu)$ is nonnegative and normalized, i.e.
\begin{eqnarray}\label{eq.4}
\int w(X,\mu,\nu)dX=1.
\end{eqnarray}
This can be checked using integrations over $X$ in (\ref{eq.3})
and property of Dirac delta-function. The function $w(X,\mu,\nu)$
is called symplectic tomogram. The physical meaning of the
symplectic tomogram is the following one. If the parameters $\mu$
and $\nu$ are represented in the form $\mu=s\cos\theta$,
$\nu=s^{-1}\sin\theta$ the symplectic tomogram is the probability
distribution of the particle position measured in the reference
frame in the particle phase-space with following properties. Let
us first scale the axes of the reference frame $q\rightarrow
sq=q'$, $p\rightarrow s^{-1}p=p'$ and  than rotate new resealed
axes $q'$ and $p'$ by the angle $\theta$. Then we get reference
frame with axes where $q''=q'\cos\theta+p'\sin\theta$ and
$q''=-q'\sin\theta+p'\cos\theta$. The coordinate $X$ it the
particle position measured in the reference frame with new
coordinates $q''$ and $p''$. Thus taking $\mu=s\cos\theta$,
$\nu=s^{-1}\sin\theta$ we have
\begin{eqnarray}\label{eq.5}
X=\mu q + \nu p.
\end{eqnarray}
The Radon transform (\ref{eq.3}) has inverse, i.e.
\begin{eqnarray}\label{eq.6}
f(q,p)=\frac{1}{4\pi^2}\int w(X,\mu,\nu) e^{i(X-\mu q-\nu
p)}dXd\mu d\nu.
\end{eqnarray}
The symplectic tomogram has the homogeneity property
\begin{eqnarray}\label{eq.7}
w(\lambda
X,\lambda\mu,\lambda\nu)=\frac{1}{|\lambda|}w(X,\mu,\nu).
\end{eqnarray}
This property follows from homogeneity property of Dirac
delta-function $\delta(\lambda y)=\frac{1}{|\lambda|}\delta(y)$.
For many particles one has the multidimensional Radon transform
\begin{eqnarray}\label{eq.8}
w(\vec{X},\vec{\mu},\vec{\nu})=\int
f(\vec{q},\vec{p})\prod_{k=1}^N\delta(X_k-\mu_k q_k-\nu_k
p_k)d\vec{q}d\vec{p}.
\end{eqnarray}
This symplectic tomogram also determines the probability
distribution $f(\vec{q},\vec{p})$ on the phase-space of the
system, i.e.
\begin{eqnarray}\label{eq.9}
f(\vec{q},\vec{p})=\frac{1}{(4\pi^2)^N}\int
w(\vec{X},\vec{\mu},\vec{\nu}) (\prod_{k=1}^N e^{i(X_k-\mu_k
q_k-\nu_k p_k)})d\vec{X}d\vec{\mu}d\vec{\nu}.
\end{eqnarray}

The symplectic tomogram (\ref{eq.8}) is nonnegative and
normalized, i.e.
\begin{eqnarray}\label{eq.10}
w(\vec{X},\vec{\mu},\vec{\nu})\geq 0
\end{eqnarray}
and
\begin{eqnarray}\label{eq.11}
\int w(\vec{X},\vec{\mu},\vec{\nu})d\vec{X}=1.
\end{eqnarray}
The physical meaning of the symplectic tomogram
$w(\vec{X},\vec{\mu},\vec{\nu})$ is the following one. It is the
joint probability distribution function of $N$ random variables
(positions) $X_k$ measured in specific reference frame in the
system phase-space subjected to operation of rescaling
$q_k\rightarrow s_kq_k$, $p_k\rightarrow s_k^{-1}p_k$ and then
rotated by the angle $\theta_k$. Then the position $X_k$ is
measured in the reference frame where all the axes are transformed
using this prescription. Important property which follows from the
physical meaning of the tomogram $w(\vec{X},\vec{\mu},\vec{\nu})$
is the reduction property, i.e.
\begin{eqnarray}\label{eq.12}
\int
w(\vec{X},\vec{\mu},\vec{\nu})dX_N=\tilde{w}(\vec{X}',\vec{\mu}',\vec{\nu}').
\end{eqnarray}
Here
$\vec{X}'=(X_1,X_2,...,X_{N-1})$,$\vec{\mu}'=(\mu_1,\mu_2,...,\mu_{N-1})$
and $\vec{\nu}'=(\nu_1,\nu_2,...,\nu_{N-1})$. This reduction
property corresponds to the reduction property of the probability
distribution function
\begin{eqnarray}\label{eq.13}
\int f(\vec{q},\vec{p})dq_Ndp_N=\tilde{f}(\vec{q}',\vec{p}')
\end{eqnarray}
where $\vec{q}'=(q,...,q_{N-1})$, $\vec{p}'=(p,...,p_{N-1})$. The
function $\tilde{f}(\vec{q}',\vec{p}')$ is the probability
distribution for the subsystem of $N-1$ particles. This
probability distribution determines the symplectic tomogram
$\tilde{w}(\vec{X}',\vec{\mu}',\vec{\nu}')$ by using the Radon
transform
\begin{eqnarray}\label{eq.14}
\int f(\vec{q}',\vec{p}')\prod_{k=1}^{N-1}\delta(X_k-\mu_k
q_k-\nu_k
p_k)d\vec{q}d\vec{p}=\tilde{w}(\vec{X}',\vec{\mu}',\vec{\nu}').
\end{eqnarray}
The symplectic tomogram of multidimensional system has the
homogeneity property
\begin{eqnarray}\label{eq.15}
w(\lambda_1X_1,\lambda_2X_2,...,\lambda_NX_N,
\lambda_1\mu_1,\lambda_2\mu_2,...,\lambda_N\mu_N,\lambda_1\nu_1,
\lambda_2\nu_2,...,\lambda_N\nu_N)=(\prod_{k=1}^N\frac{1}{|\lambda_k|})w(\vec{X},\vec{\mu},\vec{\nu}).
\end{eqnarray}
The function must satisfy the condition
\begin{eqnarray}\label{eq.16}
\int w(\vec{X},\vec{\mu},\vec{\nu})\prod_{k=1}^N e^{i(X_k-\mu_k
q_k-\nu_k p_k)})d\vec{X}d\vec{\mu}d\vec{\nu}\geq 0
\end{eqnarray}
since this integral determines the probability distribution of the
system on the phase-space.

\section{Liouville equation in tomographic representation}
The simplest kinetic equation in classical statistical mechanics
is Liouville equation for the probability distribution $f(q,p,t)$
of one particle on the phase-space. This equation reads
\begin{eqnarray}\label{eq.2.1}
\frac{\partial f(q,p,t)}{\partial t}+p\frac{\partial
f(q,p,t)}{\partial q}-\frac{\partial U(q)}{\partial
q}\frac{\partial f(q,p,t)}{\partial p}=0.
\end{eqnarray}
Here $U(q)$ is potential energy and this equation corresponds to
the Newton equation of motion
\begin{eqnarray}\label{eq.2.2}
\dot{p}=-\frac{\partial U(q)}{\partial q}.
\end{eqnarray}
In the case of many degrees of freedom the Hamiltonian reads
\begin{eqnarray}\label{eq.2.3}
H(q_1,q_2,...,q_N,p_1,p_2,...,p_N)=\sum_{j=1}^N\frac{p_j^2}{2}+U(q_1,q_2,...,q_N).
\end{eqnarray}
The Newton equation of motion corresponding to this Hamiltonian
has the form
\begin{eqnarray}\label{eq.2.4}
\dot{p_j}=-\frac{\partial U(q_1,...,q_N)}{\partial q_j}.
\end{eqnarray}
The Liouville kinetic equation in this case reads
\begin{eqnarray}
&&\frac{\partial f(q_1,q_2,...,q_N,p_1,p_2,...,p_N,t)}{\partial
t}+\sum_{j=1}^Np_j\frac{\partial
f(q_1,q_2,...,q_N,p_1,p_2,...,p_N,t)}{\partial
q_j}\nonumber\\
&&-\sum_{j=1}^N\frac{\partial
f(q_1,q_2,...,q_N,p_1,p_2,...,p_N,t)}{\partial p_j}\frac{\partial
U(q_1,q_2,...,q_N,p_1,p_2,...,p_N)}{\partial q_j}=0.\label{eq.2.5}
\end{eqnarray}
The Liouville kinetic equation can be rewritten in tomographic
probability representation. It means that we replace the
probability density $f(q_1,q_2,...,q_N,p_1,p_2,...,p_N,t)$ by its
Radon transform and write down equation for the tomographic
probability density. For one degree of freedom we use the
correspondence rules
\begin{eqnarray}
&&f(q,p,t)\longleftrightarrow w(X,\mu,\nu,t);
\nonumber\\
&&pf(q,p,t)\longleftrightarrow -(\frac{\partial}{\partial
X})^{-1}\frac{\partial}{\partial \nu}w(X,\mu,\nu,t);
\nonumber\\
&&\frac{\partial}{\partial q}f(q,p,t)\longleftrightarrow
\mu\frac{\partial}{\partial X}w(X,\mu,\nu,t);
\nonumber\\
&&qf(q,p,t)\longleftrightarrow -(\frac{\partial}{\partial
X})^{-1}\frac{\partial}{\partial \mu}w(X,\mu,\nu,t);
\nonumber\\
&&\frac{\partial}{\partial p}f(q,p,t)\longleftrightarrow
\nu\frac{\partial}{\partial X}w(X,\mu,\nu,t).\label{eq.2.4a}
\end{eqnarray}
Then we get the Liouville kinetic equation in the tomographic
probability representation
\begin{eqnarray}
&&\frac{\partial w(X,\mu,\nu,t)}{\partial t}-\mu\frac{\partial
w(X,\mu,\nu,t)}{\partial \nu}\nonumber\\- &&\frac{\partial
U}{\partial q}(q\rightarrow -(\frac{\partial}{\partial
X})^{-1}\frac{\partial}{\partial \mu})\nu\frac{\partial
w(X,\mu,\nu,t)}{\partial X}=0.\label{eq.2.5a}
\end{eqnarray}
For $N$ degrees of freedom the correspondence rules have the form
\begin{eqnarray}
&&f(q_1,q_2,...,q_N,p_1,p_2,...,p_N,t)\longleftrightarrow
w(X_1,X_2,...,X_N,\mu_1,\mu_2,...,\mu_N,\nu_1,\nu_2,...,\nu_N,t);
\nonumber\\
&&p_jf(q_1,q_2,...,q_N,p_1,p_2,...,p_N,t)\longleftrightarrow
-(\frac{\partial}{\partial X_j})^{-1}\frac{\partial}{\partial
\nu_j}w(X_1,X_2,...,X_N,\mu_1,\mu_2,...,\mu_N,\nu_1,\nu_2,...,\nu_N,t);
\nonumber\\
&&\frac{\partial}{\partial
q_j}f(q_1,q_2,...,q_N,p_1,p_2,...,p_N,t)\longleftrightarrow
\mu_j\frac{\partial}{\partial
X_j}w(X_1,X_2,...,X_N,\mu_1,\mu_2,...,\mu_N,\nu_1,\nu_2,...,\nu_N,t);
\nonumber\\
&&q_jf(q_1,q_2,...,q_N,p_1,p_2,...,p_N,t)\longleftrightarrow
-(\frac{\partial}{\partial X_j})^{-1}\frac{\partial}{\partial
\mu_j}w(X_1,X_2,...,X_N,\mu_1,\mu_2,...,\mu_N,\nu_1,\nu_2,...,\nu_N,t);
\nonumber\\
&&\frac{\partial}{\partial
p_j}f(q_1,q_2,...,q_N,p_1,p_2,...,p_N,t)\longleftrightarrow
\nu_j\frac{\partial}{\partial
X_j}w(X_1,X_2,...,X_N,\mu_1,\mu_2,...,\mu_N,\nu_1,\nu_2,...,\nu_N,t).\label{eq.2.6}
\end{eqnarray}
The operator $(\frac{\partial}{\partial X})^{-1}$ acting on a
function $\varphi(X)$ gives the function $\Phi(X)$ satisfying the
equality
\begin{eqnarray}\label{eq.2.7}
\frac{d\Phi(X)}{dX}=\varphi(X).
\end{eqnarray}
In case of representing the functions $\varphi(X)$ and $\Phi(X)$
in form of Fourier integrals, i.e.
\begin{eqnarray}\label{eq.2.8}
\varphi(X)=\int\tilde{\varphi}(k)e^{ikX}dk
\end{eqnarray}
and
\begin{eqnarray}\label{eq.2.9}
\Phi(X)=\int\tilde{\phi}(k)e^{ikX}dk
\end{eqnarray}
we define the action of the operator $(\frac{\partial}{\partial
X})^{-1}$ on the function $\varphi(X)$ by equality
\begin{eqnarray}\label{eq.2.10}
(\frac{\partial}{\partial X})^{-1}\varphi(X)=\int
\frac{1}{ik}\tilde{\phi}(k)e^{ikX}dk
\end{eqnarray}
which removes the ambiguity in choice of constant in the function
$\Phi(X)$. Now we can write down the system with many degrees of
freedom in the tomographic probability representation. To do this
we make Radon transform in (\ref{eq.2.5}). Then we get
\begin{eqnarray}
&&\frac{\partial w(\vec{X},\vec{\mu},\vec{\nu},t)}{\partial
t}-\sum_{j=1}^N\mu_j\frac{\partial
w(\vec{X},\vec{\mu},\vec{\nu},t)}{\partial \nu_j}-\nonumber\\
&&-\sum_{j=1}^N\frac{\partial U}{\partial q_j}(q_1\rightarrow
-(\frac{\partial}{\partial X_1})^{-1}\frac{\partial}{\partial
\mu_1},q_2\rightarrow -(\frac{\partial}{\partial
X_2})^{-1}\frac{\partial}{\partial \mu_2},...,\nonumber\\
&&q_N\rightarrow -(\frac{\partial}{\partial
X_N})^{-1}\frac{\partial}{\partial \mu_N})\nu_j\frac{\partial
w(\vec{X},\vec{\mu},\vec{\nu},t)}{\partial X_j}=0.\label{eq.2.11}
\end{eqnarray}
The written equation describes evolution of tomographic
probability distribution function which corresponds to standard
Liouville equation for the probability density on the phase-space.

\section{Reduced Liouville equation}
The Bogolyubov chain of equations for probability distribution
function of one particle can be obtained using Liouville equation
for many particles. The ansatz is the following. One integrates
the Liouville equation over $N-1$ pairs of variables
$q_2,q_3,...,q_N,p_2,p_3,...,p_N$. Then one has the equation for
one degree of freedom only with the contribution of the
interaction of one particle with the rest. We provide this
derivation considering system of two particles with the Liouville
equation of the form

\begin{eqnarray}
&& \frac{\partial}{\partial t}
f(q_1,q_2,p_1,p_2,t)+p_1\frac{\partial
f(q_1,q_2,p_1,p_2,t)}{\partial q_1}+p_2\frac{\partial
f(q_1,q_2,p_1,p_2,t)}{\partial q_2}-\nonumber\\
&& -\frac{\partial U(q_1,q_2)}{\partial q_1}\frac{\partial
f(q_1,q_2,p_1,p_2,t)}{\partial p_1}-\frac{\partial
U(q_1,q_2)}{\partial q_2}\frac{\partial
f(q_1,q_2,p_1,p_2,t)}{\partial p_2}=0.\label{eq.2.12}
\end{eqnarray}

We integrate this equation over variables of second particle
$q_2,p_2$. Let us assume that potential energy depends on distance
between the particles, i.e.
\begin{eqnarray}
&&U(q_1,q_2)=U(|q_1-q_2|)=U(|X_{1,2}|),\nonumber\\
&&  X_{1,2}=q_1-q_2.\label{eq.2.12a}
\end{eqnarray}
Then we get

\begin{eqnarray}
&& \frac{\partial}{\partial
t}\tilde{f}(q_1,p_1,t)+p_1\frac{\partial}{\partial
q_1}\tilde{f}(q_1,p_1,t)+\int p_2\frac{\partial}{\partial
q_2}f(q_1,q_2,p_1,p_2,t)dq_2dp_2-\nonumber\\
&&-\int U'(|X_{1,2}|)(\frac{\partial}{\partial
q_1}|q_1-q_2|)\frac{\partial}{\partial
p_1}f(q_1,q_2,p_1,p_2,t)dq_2dp_2+\nonumber\\
&& +\int U'(|X_{1,2}|)(\frac{\partial}{\partial
q_1}|q_1-q_2|)\frac{\partial}{\partial
p_2}f(q_1,q_2,p_1,p_2,t)dq_2dp_2=0.\label{eq.2.13}
\end{eqnarray}
We use the notation
\begin{eqnarray}\label{eq.2.14}
\tilde{f}(q_1,p_1,t)=\int f(q_1,q_2,p_1,p_2,t)dq_2dp_2.
\end{eqnarray}
This equation can be transformed and analogous procedure can be
done with Liouville equations (\ref{eq.2.5}). Now we make the
reduction of tomographic Liouville equation (\ref{eq.2.11})on the
example of two particles. The equation (\ref{eq.2.11}) for two
particles reads
\begin{eqnarray}
&& \frac{\partial w(X_1,X_2,\mu_1,\mu_2,\nu_1,\nu_2,t)}{\partial
t}-(\mu_1\frac{\partial}{\partial\nu_1}+\mu_2\frac{\partial}{\partial\nu_2})w(X_1,X_2,\mu_1,\mu_2,\nu_1,\nu_2,t)-\nonumber\\
&&-U'(|(\frac{\partial}{\partial
X_1})^{-1}\frac{\partial}{\partial
\mu_1}-(\frac{\partial}{\partial
X_2})^{-1}\frac{\partial}{\partial
\mu_2}|)sgn[-(\frac{\partial}{\partial
X_1})^{-1}\frac{\partial}{\partial
\mu_1}+(\frac{\partial}{\partial
X_2})^{-1}\frac{\partial}{\partial \mu_2}]w(X_1,X_2,\mu_1,\mu_2,\nu_1,\nu_2,t)-\nonumber\\
&&-U'(|(\frac{\partial}{\partial
X_1})^{-1}\frac{\partial}{\partial
\mu_1}-(\frac{\partial}{\partial
X_2})^{-1}\frac{\partial}{\partial
\mu_2}|)sgn[-(\frac{\partial}{\partial
X_2})^{-1}\frac{\partial}{\partial
\mu_2}+(\frac{\partial}{\partial
X_1})^{-1}\frac{\partial}{\partial
\mu_1}]w(X_1,X_2,\mu_1,\mu_2,\nu_1,\nu_2,t)=0.\nonumber\\
&& \label{eq.2.15}
\end{eqnarray}
We can get the reduction of this equation for tomogram describing
the state of one particle integrating it over variable $X_2$. One
has
\begin{eqnarray}
&&\frac{\partial \tilde{w}(X_1,\mu_1,\nu_1,t)}{\partial
t}-\mu_1\frac{\partial}{\partial\nu_1}\tilde{w}(X_1,\mu_1,,\nu_1,t)+\nonumber\\
&& +\int dX_2[-U'(|(\frac{\partial}{\partial
X_1})^{-1}\frac{\partial}{\partial
\mu_1}-(\frac{\partial}{\partial
X_2})^{-1}\frac{\partial}{\partial
\mu_2}|)sgn[-(\frac{\partial}{\partial
X_1})^{-1}\frac{\partial}{\partial
\mu_1}+(\frac{\partial}{\partial
X_2})^{-1}\frac{\partial}{\partial \mu_2}]w(X_1,X_2,\mu_1,\mu_2,\nu_1,\nu_2,t)-\nonumber\\
&&-U'(|(\frac{\partial}{\partial
X_1})^{-1}\frac{\partial}{\partial
\mu_1}-(\frac{\partial}{\partial
X_2})^{-1}\frac{\partial}{\partial
\mu_2}|)sgn[-(\frac{\partial}{\partial
X_2})^{-1}\frac{\partial}{\partial
\mu_2}+(\frac{\partial}{\partial
X_1})^{-1}\frac{\partial}{\partial
\mu_1}]w(X_1,X_2,\mu_1,\mu_2,\nu_1,\nu_2,t)]=0.\label{eq.2.16}
\end{eqnarray}
The obtained equation (\ref{eq.2.16}) is consistent with taking
Radon transform of the reduced Liouville equation (\ref{eq.2.13}).
The Bogolyubov chain of equations in case of many particles can be
obtained by analogous procedure applied to equation
(\ref{eq.2.11}). In this case we integrate the equation over
variables $X_2,X_3,...,X_N$.

\section{Conclusion}
We point out the main results of the work. We considered Liouville
equation for $N$ particles and using the integral Radon transform
of the probability density on phase-space rewrote the equation in
the tomographic probability representation. The Bogolyubov chain
of the equations obtained by averaging the Liouville equations
over subsystem degrees of freedom is also written in the
tomographic probability representation. We hope to extend in
future publications the tomographic representation analysis to the
quantum kinetic equations using the Bogolyubov chain approach to
Moyal equation \cite{Moy49} for Wigner function.

\section{Acknowledgement}
V.N.C. and V.I.M. thank Russian Foundation for Basic Research for
the support under Project Nos.07-02-00598,08-02-90300, and
09-02-00142.

\end{document}